\documentclass[letter]{jpsj2} 
\def\runauthor{Hiroyuki {\sc Takagiwa} {\it et al.}}

\title{Correlation between Superconducting Carrier Density and
Transition Temperature in NbB$_{2+x}$}
\author{
Hiroyuki {\sc Takagiwa}, Sogo {\sc Kuroiwa}, Maki {\sc Yamazawa}, 
Jun {\sc Akimitsu},\\
Kazuki {\sc Ohishi}$^1$, Akihiro {\sc Koda}$^1$, 
Wataru {\sc Higemoto}$^1$\thanks{Present Address: Japan Atomic Energy Research Institure (JAERI), Tokai, Ibaraki 319-1195} 
and Ryosuke {\sc Kadono}$^1$\thanks{Also at the Graduate University for Advanced Studies (SOKENDAI)} 
}
\inst{
Department of Physics, Aoyama-Gakuin University, Sagamihara, Kanagawa 229-8558\\
$^1$Institute of Materials Structure Science, High Energy Accelerator Research Organization (KEK), Tsukuba, Ibaraki 305-0801\\
}

\recdate{\today}
\abst{We report on the magnetic penetration depth, $\lambda$, in a type II 
superconductor NbB$_{2+x}$ determined by muon spin rotation method. We show 
in the sample with $x=0.1$ that $\lambda$ at 2.0 K is independent of an applied magnetic field. 
This suggests that the superconducting order parameter in NbB$_{2+x}$ is 
isotropic, as expected for 
conventional BCS superconductors. Meanwhile, the superconducting carrier density   
($\propto\lambda^{-2}$) exhibits an interesting tendency of increase with increasing
$T_{\rm c}$ (where $T_{\rm c}$ varies with $x$). Possible origin of such behavior
is discussed in comparison with the case of exotic superconductors. }

\kword{NbB$_{2+x}$, superconductivity, $\mu$SR, magnetic penetration depth, 
carrier density}

\begin{document}
\sloppy
\maketitle
The recent revelation of 
superconductivity in MgB$_2$ with high transition temperature ($T_{\rm c}$  
= 39K)\cite{Nagamatsu:01} has revived much interest in
hexagonal diborides which were once subjected to active studies from 1950's through 
70's\cite{Ziegler:53, Cooper:70, Leyarovska:79}.  The discovery of MgB$_2$ 
prompted various theoretical and experimental investigations to clarify the origin of 
the high 
$T_{\rm c}$ seemingly exceeding the so-called `BCS-barrier' ($\sim30$ K). 
These compounds have a simple layered structure along $c$-axis, 
consisting of an alternative stack 
of metal ion layers forming a triangular lattice and boron atom layers 
in a honeycomb network. 
In MgB$_2$, both theoretical and experimental studies suggested that 
the strong electron-phonon interaction in accordance with the light mass of the 
organized atoms and two-dimensional character of boron atom network 
are closely related to the relatively high-$T_{\rm c}$ superconductivity. 
Moreover, it turned out that some of the unconventional properties in the 
superconducting phase can be
attributed to the multiband superconductivity, where the presence
of two energy gaps corresponding to $p_{\pi}$ and $p_{\sigma}$ orbits 
of boron atoms have been confirmed by experimental\cite{Souma:03}
and theoretical studies\cite{Liu:01}.

Among various diborides, NbB$_2$ is intriguing as it exhibits strong sensitivity
of superconducting transition temperature to the stoichiometric imperfection; 
Cooper \textit{et al.} reported $T_{\rm c}$ = 3.87 K 
for boron-rich NbB$_2$\cite{Cooper:70}, while Leyarovska \textit{et al.}  reported 
much lower $T_{\rm c}$ ( = 0.62 K) for a stoichiometric NbB$_2$\cite{Leyarovska:79}. 
Schirber \textit{et al.}  showed $T_{\rm c}$ = 9.4 K for single-crystalline sample of 
NbB$_x$ ($x \sim$ 2)\cite{Schirber:92}. Recently, Yamamoto $et~al$. reported that 
$T_{\rm c}$ changes between 2 K and 9.2 K in the samples prepared under high pressure,  probably because of the shift in the composition ratio between 
Nb and boron\cite{Yamamoto:02}.
It is noticeable that Escamilla \textit{et al.} found a maximum 
$T_{\rm c}$ = 9.8 K\cite{Escamilla:04}, which is slightly higher than that of 
elemental Nb (=9.23 K).
Thus, it might be possible to obtain a guide for ever higher $T_{\rm c}$ in hexagonal 
diborides by studying
the relationship between $T_{\rm c}$ and various parameters including chemical
composition and crystal structure in NbB$_{2+x}$.

In this paper, we report on the magnetic penetration depth, $\lambda$, 
in the mixed state of NbB$_{2+x}$ studied by muon spin rotation 
and relaxation method ($\mu$SR). The magnetic penetration depth, which reflects 
superconducting carrier density, $n_{\rm s}$ ($\propto\lambda^{-2}$), 
is determined microscopically by $\mu$SR signals 
due to the inhomogeneity of magnetic field distribution in the flux line lattice (FLL). 
We show that $\lambda$ is independent of the applied field, strongly suggesting that 
NbB$_{2+x}$ has an isotropic superconducting order parameter. 
More interestingly, a quasi-linear relationship has been observed between the muon spin 
relaxation rate $\sigma$ ($T$ = 0 K, which is proportional to $n_{\rm s}$) and 
$T_{\rm c}$ as it varies with $x$.  This result suggests that the change of $T_{\rm c}$ in NbB$_{2+x}$ depends on the superconducting carrier density.

Polycrystalline samples of NbB$_{2+x}$ ($x$ = 0, 0.03, 0.1 where $x$ is a nominal value) used in this experiment were 
synthesized from Nb powder (99.96 \%) and boron crystalline powder (99 \%). 
The pellets of these mixtures were sealed by quartz tube at vacuum pressure 
($<$ 3.0 $\times$ 10$^{-5}$ Torr) and heated at 1000 $^{\circ}$C  for 30 hours. 
These samples were confirmed to be single phase, and to decrease $a$-axis and increse $c$-axis with increasing $x$ by the X-ray powder diffraction pattern 
using the conventional X-ray powder spectrometer (RAD-C; RIGAKU). The magnetic 
susceptibility and electrical resistivity were measured with the SQUID magnetometer 
(MPMS2; Quantum Design Co., Ltd.) and the PPMS system (Quantum Design Co., Ltd.).

Figure 1 shows the temperature dependence of magnetic susceptibility upon field 
cooling at 10 Oe, where one can observe significant reduction of the susceptibility
at a well-defined temperature, $T_{\rm c}$, for respective compositions.
This is obviously due to the Meissener diamagnetism  associated with the superconducting
transition.  It is also clear in Fig.~1 that $T_{\rm c}$ increases with increasing $x$. 
The superconducting volume fraction at 1.8 K is estimated to be approximately 44, 34 and 29 \% for NbB$_2$, NbB$_{2.03}$ and NbB$_{2.1}$. The corresponding upper 
critical field ($H_{\rm c2}$(0)) are 6.4, 7.6 and 10.5 kOe respectively, which were 
estimated from the results of electrical resistivity measurements under an applied 
magnetic field.

The $\mu$SR experiment was performed on the M15 and M20 beamlines at TRIUMF which 
provides a muon beam with the momentum of 29 MeV/c. The polycrystalline samples were
mounted on sample holders having a dimension of 7 $\times$ 7 mm$^2$ (M15) or 
25 $\times$ 25 mm$^2$ (M20). The $\mu$SR spectra were obtained at $H$ = 1 kOe and at $T$ = 2 K to map out the temperature and magnetic field  dependence, respectively.
Care was taken to cool down the sample in the field-cooled condition at each applied field
to avoid the effect of flux pinning.

Since the muons stop randomly on the FLL, the muon spin precession signal 
$\hat{P}(t)$ provides a random sampling of the internal field distribution $B(\hat{r})$,
\begin{eqnarray}
\hat{P}(t) &\equiv& P_x(t) + iP_y(t) = \int_{-\infty}^{\infty}n(B){\rm exp}(i\gamma _{\mu}Bt)dB,\\
n(B) &=& \langle \delta(B(\hat{r}) - B)\rangle _r,
\end{eqnarray}
where $\gamma _{\mu}$ is the muon gyromagnetic ratio (= 2$\pi \times$ 13.553 
MHz/kOe), and $n(B)$ is the spectral density for muon precession determined by 
the local field distribution. These equations indicate that the real amplitude of the 
Fourier transformed muon precession signal corresponds to the local field distribution 
$n(B)$. The London penetration depth in the FLL state is related to the second 
moment $\langle (\Delta B)^2 \rangle$ = $\langle (B(\hat{r}) - H)^2 \rangle$ of 
the field distribution reflected in the $\mu$SR line shape\cite{Brandt:88}. In polycrystalline samples, a Gaussian distribution of local fields is a good approximation,
\begin{eqnarray}
\hat{P}(t) &\simeq& {\rm exp}(-\sigma ^2t^2/2){\rm exp}(i\gamma _{\mu}Ht),\\
\sigma &=& \gamma _{\mu} \sqrt{\langle (\Delta B)^2 \rangle}.
\end{eqnarray}
For the ideal triangular FLL with isotropic effective carrier mass $m^{\ast}$, $\lambda$ is given by the relation\cite{Brandt:88, Pincus:64, Aeppli:87},
\begin{equation}
\sigma~[\mu {\rm s}^{-1}] = 4.83 \times 10^4(1-h)[1 + 3.9(1-h)^2]^{1/2}\lambda ^{-2}~[{\rm nm}],\label{sgmv}
\end{equation}
where $ h = H/H_{{\rm c}2}$, and $\lambda$ is related to the superconducting carrier density,
\begin{equation}
\lambda ^2 = \frac{m^{\ast}c^2}{4\pi n_{\rm s}e^2},\label{lmdns}
\end{equation}
indicating that $\lambda$ is enhanced upon the reduction of $n_{\rm s}$ due to the quasiparticle excitations.

Figure 2 shows typical muon spin precession signals in NbB$_{2.1}$ 
at various temperatures under a transverse field of 1 kOe, where one can observe
the change in the lineshape due to FLL formation as the temperature
passes through $T_{\rm c}$.  The damping of precession amplitude above 
$T_{\rm c}$ is mainly due to static random local fields from $^{93}$Nb and 
$^{10, 11}$B nuclear magnetic moments. Upon cooling below $T_{\rm c}$, the spectrum 
exhibits enhanced depolarization, which indicates that magnetic field distribution becomes 
inhomogeneous due to the formation of magnetic vortices. 
In analyzing these spectra, we adopted the following fitting function,
\begin{equation}
A\hat{P}(t) = \sum_{j=1}^2 A_j\exp\left( -\frac{\sigma _j^2t^2}{2} \right)\cos(-\gamma_{\mu}B_jt + \phi _j),
\end{equation}
where the index $j$ denotes the components of normal ($j$ = 1) and superconducting 
($j$ = 2) domains, $A$ is the total positron decay asymmetry, $A_j$ is the partial 
asymmetry, $\sigma _j$ is the muon spin relaxation rate, $B_j$ is the central frequency 
and $\phi _j$ is the initial phase for respective components. We define $\sigma _1$ to 
reflect the spin relaxation in the normal state where it is dominated by the nuclear magnetic moments, $\sigma _{\rm n}$, 
and $\sigma _2$ in the superconducting state where $\sigma _2^2 = 
\sigma _{\rm n}^2 + \sigma _{\rm v}^2$ with $\sigma _{\rm v}$ being due to the 
formation of FLL. The temperature dependence of $\sigma _{\rm v}$ in NbB$_{2.1}$ 
at $H$ = 1 kOe is shown in Fig.~3.  In the FLL state, $\sigma _{\rm v}$ increases 
with decreasing temperature below $T_{\rm c}$(1 kOe) $\sim5$ K. 
According to the empirical two-fluid model, which is supposed to be 
a good approximation for the BCS theory, the following relation is expected,
\begin{equation}
\lambda (T) = \lambda (0)\frac{1}{\sqrt{1-\tau ^4}},
\end{equation}
which leads to
\begin{equation}
\sigma _{\rm v}(T) = \sigma _{\rm v}(0)(1-\tau ^4),
\end{equation}
where $\tau \equiv T/T_{\rm c}$(1 kOe). As clearly seen in Fig.~3, the agreement
between the above relation and our data is far from satisfactory. A better agreement 
is attained when we perform the fitting analysis by a similar formula,
\begin{equation}
\sigma _{\rm v}(T) = \sigma _{\rm v}(0)(1-\tau ^2).
\end{equation}
It yields $T_{\rm c}$ = 5.3(1) K when $T_{\rm c}$ is a free parameter, which is in good
agreement with the value estimated from the electrical resistivity measurement under 
the same field ($\sim 5.0$ K). We also performed fitting analysis by a weak-coupling 
BCS model (w-BCS)\cite{Muhlschlegel:59}, which turned out to reproduce the 
experimental data very well. Thus, these analyses indicate the difficulty to
assess the structure of order parameter in NbB$_{2+x}$ based solely on the temperature
dependence of $\sigma_{\rm v}$ with the data above 2 K.  

The situation is much improved by looking into the magnetic 
field dependence of $\sigma_{\rm v}$. 
As shown in Fig.~4(a)  $\sigma _{\rm v}$ in NbB$_{2.1}$ decreases with increasing 
external field, where the solid line represents the fitting result by eq.(\ref{sgmv}) with 
$H_{\rm c2}$ = 6.7 kOe determined by electrical resistivity measurements. 
The corresponding field dependence of $\lambda$ is shown in Fig.~4(b), where
 $\lambda$ is mostly independent of applied magnetic field.   For a quantitative evaluation
on the strength of the pair-breaking effect by external field, 
we performed a fitting analysis by the following simple linear relation,
\begin{equation}
\lambda (h) = \lambda (0) [1+\eta h],\label{lmdh}
\end{equation}
where $\eta$ is a dimension-less parameter which represents the strength of pair 
breaking effect, and $h \equiv H/H_{\rm c2}$(2 K) with $H_{\rm c2}$(2 K) = 6.7 kOe. 
From the fitting analysis for the data by eq.(\ref{lmdh}), we obtain $\eta$ = 0.02(2) with 
$\lambda(0)$ = 233(2) nm, where the result is shown as a solid line in Fig.~4(b). 
It has been established experimentally that the parameter $\eta$ represents the degree
of anisotropy in the superconducting order parameter\cite{Kadono:04}, since the
pair-breaking mechanism due to the quasi-classical Doppler shift has come to 
wide recognition\cite{Volovik:93}.
 For example, in Y(Ni$_{0.8}$Pt$_{0.2}$)B$_2$C\cite{Ohishi:03}, 
Cd$_2$Re$_2$O$_7$\cite{Kadono:02} and V$_3$Si ($h<0.5$)\cite{Sonier:04}, 
which behave as conventional BCS superconductors with isotropic $s$-wave pairing, 
$\eta$ is reported to be nearly zero. On the other hand, $\eta$ = 5 -- 6.6 
in a high-$T_{\rm c}$ cuprate YBa$_2$Cu$_3$O$_{7-\delta}$ 
which has a $d$-wave pairing\cite{Sonier:97}. 
In the case of YNi$_2$B$_2$C which turned out to be an anisotropic
 $s$-wave superconductor, $\eta \simeq 1$ is reported\cite{Ohishi:02}. 
From the comparison of these results with our result, it is strongly suggested that 
NbB$_{\rm 2.1}$ has an isotropic $s$-wave order parameter.

We made additional measurements for  NbB$_{2+x}$ with $x=$ 0 and 0.03 to 
clarify the relation between the bulk
$T_{\rm c}$ and microscopic parameters such as $\lambda$. 
The relation between $T_{\rm c}$ and $\sigma _{\rm v} (0)$, which is
extrapolated from the temperature dependence of $\sigma _{\rm v} (T)$, 
is shown in Fig.~5. 
By the comparison between eqs.(\ref{sgmv}) and (\ref{lmdns}), 
the following relation is derived;  
\begin{equation}
\sigma _{\rm v} \propto \frac{1}{\lambda ^2} \propto \frac{n_{\rm s}}{m^{\ast}}.
\end{equation}
It is inferred from Fig.~5 that $T_{\rm c}$ increases with increasing 
$n_{\rm s}/m^{\ast}$ with a quasi-linear relation ($T_{\rm c} 
\propto n_{\rm s}/m^{\ast}$) . This strongly suggests that the change 
of $T_{\rm c}$ is associated with that of the superconducting carrier density.
It is interesting to note that this behavior has a certain similarity with the case of 
exotic (e.g. high-$T_{\rm c}$ cuprate, organic, heavy-fermion) superconductors\cite{Uemura:91}.  According to the MEM/Rietveld analysis on
the synchrotron radiation X-ray measurements, this corresponds to the change in 
the number of electrons at Nb- and B-sites\cite{Takagiwa:05}.
It is generally expected from the BCS theory that 
\begin{equation}
T_{\rm c} \propto \exp(-1/N(0)V),
\end{equation}
where $N(0)$ is the density of state at the Fermi surface and $V$ is the electron-phonon
coupling constant. Assuming that $V$ is independent of the slight change in the chemical
composition, our result suggests that $N(0)$ is strongly dependent of the 
boron content $x$ in NbB$_{2+x}$.  This might suggest that a similar effect may have
to be considered in understanding the Uemura-plot for exotic superconductors\cite{Uemura:91}.
However, we note that the possibility of  $x$-dependent electron-phonon coupling
cannot be ruled out solely based on the present result, considering the possible modulation
of lattice structure by non-stoichiometric boron contents.

In summary, we have investigated the magnetic penetration depth in the FLL state of 
NbB$_{2+x}$ using $\mu$SR method. The field dependence of $\lambda$ strongly 
suggests that the superconducting order parameter in NbB$_{2+x}$ is isotropic 
and thereby described by the BCS $s$-wave pairing.
The composition dependence of $T_{\rm c}$ in this compound 
suggests that the superconducting carrier density is a limiting factor in determining
$T_{\rm c}$ as inferred from the relationship between $T_{\rm c}$ and 
$\sigma _{\rm v}$.

We would like to thank TRIUMF stuff for their technical support during experiment. 
This work was partially supported by a Grant-in-Aid for Scientific Research on Priority 
Areas from the Ministry of Education, Culture, Sports, Science and Technology of Japan.

\newpage
\begin{figure}[b]
\begin{center} \includegraphics[width=0.8\linewidth]{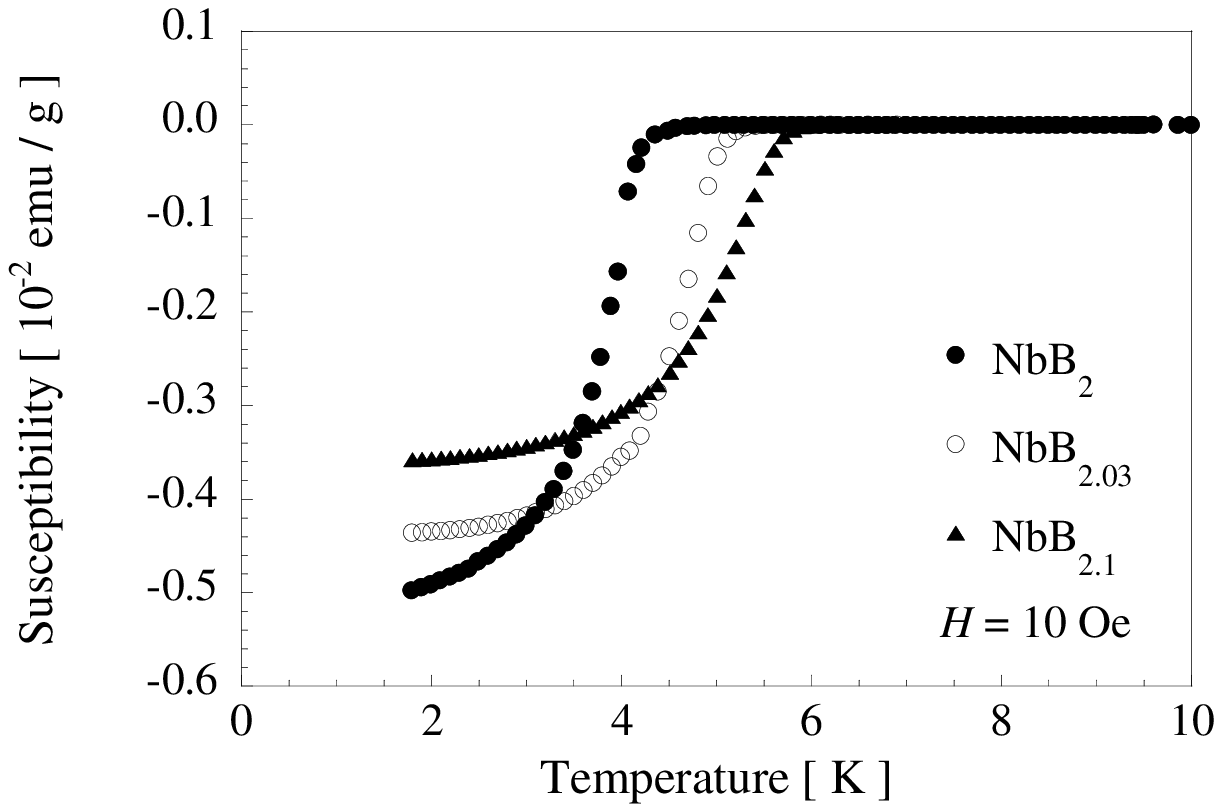}
\caption{Temperature dependence of the magnetic susceptibility at 10 Oe
measured under field-cooling condition.} \label{fig1}
\end{center}
\end{figure}

\begin{figure}[t]
\begin{center} \includegraphics[width=0.8\linewidth]{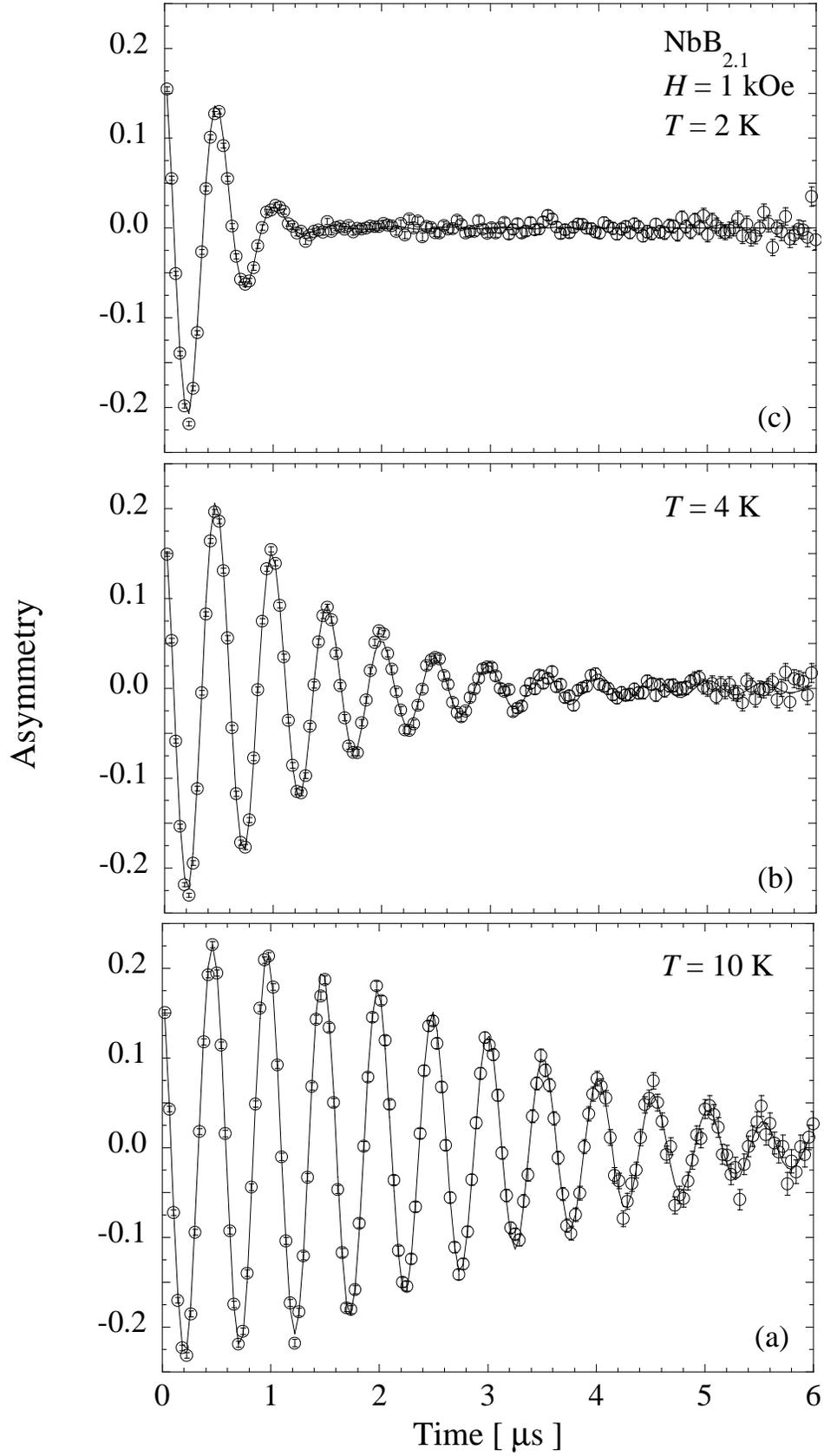}
\caption{Typical muon spin precession signals in NbB$_{2.1}$ ($T_{\rm c}$ = 5.0 K) under a transverse field 
of 1 kOe at temperatures of (a) 10 K (above $T_{\rm c}$), (b) 4 K and (c)  2 K 
(below $T_{\rm c}$).} \label{fig2}
\end{center}
\end{figure}

\begin{figure}[t]
\begin{center} \includegraphics[width=0.8\linewidth]{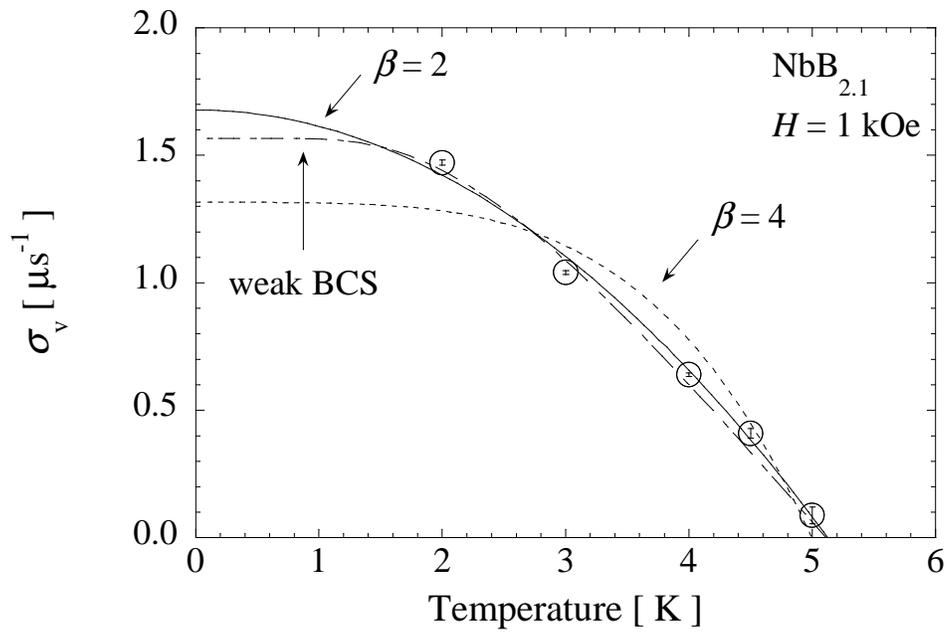}
\caption{Temperature dependence of the muon spin relaxation rate $\sigma _{\rm v}$ 
due to flux line lattice formation at $H$ = 1 kOe. The curves are results of fitting by a 
relation $\sigma _{\rm v}(T) = \sigma (0)[1-(T/T_{\rm c})^{\beta}]$ 
and weak-coupling BCS model (dot-dashed curve). The solid curve is for $\beta$ =2 
and the dotted curve is for $\beta$ = 4.} \label{fig3}
\end{center}
\end{figure}

\begin{figure}[t]
\begin{center} \includegraphics[width=0.8\linewidth]{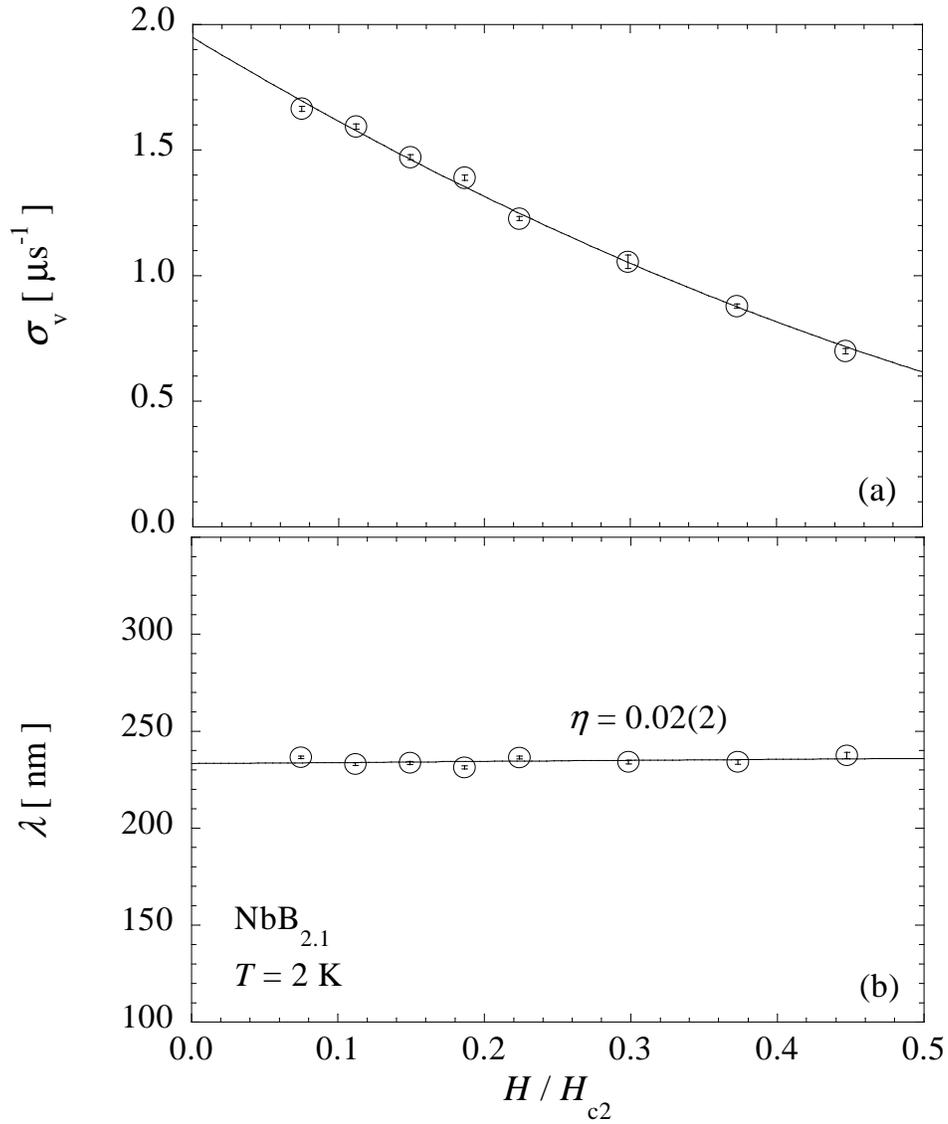}
\caption{Magnetic field dependence of (a) the muon spin relaxation rate 
$\sigma _{\rm v}$ and (b) the magnetic penetration depth $\lambda$ in NbB$_{2.1}$ 
at 2 K. The curves in (a) and (b) are results of fitting by eqs.(\ref{sgmv}) with $\lambda$ proportional to $\lambda$(0)[1 + $\eta h$] and 
(\ref{lmdh}). } \label{fig4}
\end{center}
\end{figure}

\begin{figure}[t]
\begin{center} \includegraphics[width=0.8\linewidth]{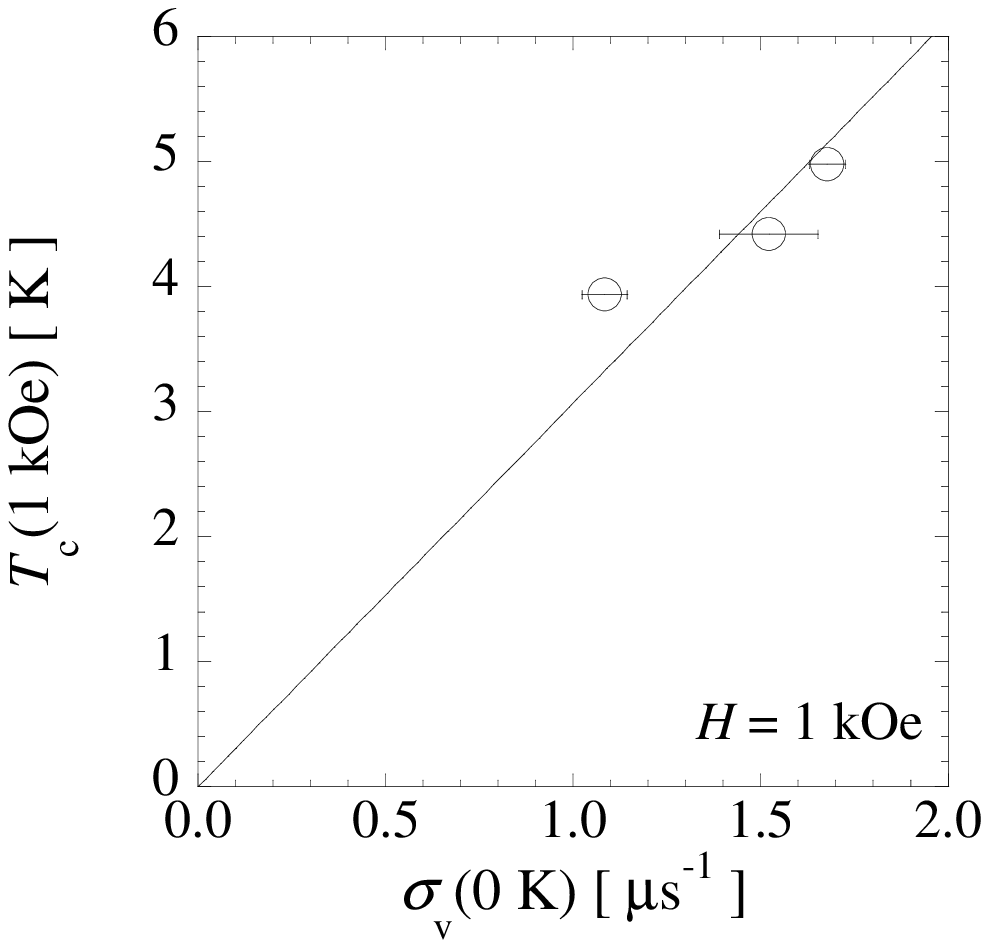}
\caption{The correlation between $T_{\rm c}$ at $H$ = 1 kOe 
and $\sigma _{\rm v}$($T$ = 0 K) in NbB$_{2+x}$, where the latter is estimated 
from the temperature dependence of $\sigma _{\rm v}$.} \label{fig5}
\end{center}
\end{figure}


\begin{thebibliography}{99}
\bibitem{Nagamatsu:01} J. Nagamatsu, N. Nakagawa, T. Muranaka, Y. Zenitani and J. Akimitsu: Nature {\bf 410} (2001) 63.
\bibitem{Ziegler:53} W. T. Ziegler and R. A. Young: Phys. Rev. {\bf 90} (1953) 115.
\bibitem{Cooper:70} A. S. Cooper, E. Corenzwit, L. D. Inginotti, B. T. Matthias and W. H. Zachariasen: Proc. Natl. Acad. Sci. USA {\bf 67} (1970) 313.
\bibitem{Leyarovska:79} L. Leyarovska and E. Layarovski: J. Less. Com. Met. {\bf 67} (1979) 249.
\bibitem{Souma:03} For example, S. Souma, Y. Machida, T. Sato, T. Takahashi, H. Matsui, S. -C. Wang, H. -B. Yang, H. Ding, A. Kaminski, J. C. Campuzano, S. Sasaki and K. Kadowaki: Nature {\bf 423} (2003) 65.
\bibitem{Liu:01} For example, Amy Y. Liu, I. I. Mazin and Jens Lortus: Phys. Rev. Lett. {\bf 87} (2001) 087005.
\bibitem{Yamamoto:02} A. Yamamoto, C. Takao, T. Masui, M. Izumi and S. Tajima: Physica C {\bf 383} (2002) 197.
\bibitem{Schirber:92} J. E. Schirber, D. L. Overmyer,B. Morosin, E. L. Venturini, R. Baughman, D. Emin, H. Klensnar and T. Sdelage: Phys. Rev. B {\bf 45} (1992) 10787.
\bibitem{Escamilla:04} R. Escamilla, O. Lovera, T. Akachi, A. Dur\'{a}n, R. Falconi, F. Morales and R. Escudero: J. Phys.: Condens. Matter {\bf 16} (2004) 5979.
\bibitem{Brandt:88} E. H. Brandt: Phys. Rev. B {\bf 37} (1988) 2349.
\bibitem{Pincus:64} P. Pincus, A. C. Gossard, V. Javvarino and J. H. Wernick: Phys. Rev. Lett. {\bf 13} (1964) 21.
\bibitem{Aeppli:87} G. Aeppli, R. J. Cava, E. J. Ansaldo, J. H. Brewer, S. R. Kreitzman, G. M. Luke, D. R. Noakes and R. F. Kiefl: Phys. Rev. B {\bf 35} (1987) 7129.
\bibitem{Muhlschlegel:59} B. M$\ddot{\rm u}$hlschlegel: Z. Phys. {\bf 155} (1959) 313.
\bibitem{Kadono:04} R. Kadono:  J. Phys.: Condens. Matt. {\bf 16} (2004) S4421.
\bibitem{Volovik:93} G.E. Volovik:  Sov. Phys. JETP Lett. {\bf 58} (1993) 469.
\bibitem{Ohishi:03} K. Ohishi, K. Kakuta, J. Akimitsu, A. Koda, W. Higemoto, R. Kadono, J. E. Sonier, A. N. Price, R. I. Miller, R. F. Kiefl, M. Nohara, H. Suzuki and H. Takagi: Physica B {\bf 326} (2003) 364.
\bibitem{Kadono:02} R. Kadono, W. Higemoto, A.Koda, Y. Kawasaki, M. Hanawa and Z. Hiroi: J. Phys. Soc. Jpn. {\bf 71} (2002) 709.
\bibitem{Sonier:04} J. E. Sonier, F. D. Callaghan, R. I. Miller, E. Boaknin, L. Taillefer, R. F. kiefl, J. H. Brewer, K. F. Poon and J. D. Brewer: Phys. Rev. Lett. {\bf 93} (2004) 017002.
\bibitem{Sonier:97} J. E. Sonier, J. H. Brewer, R. F. Kiefl, D. A. Bonn, S. R. Dunsiger, W. N. Hardy, R. Liang, W. A. MacFarlane, R. I. Miller and T. M. Riseman: Phys. Rev. Lett. {\bf 79} (1997) 2875.
\bibitem{Ohishi:02} K. Ohishi, K. Kakuta, J. Akimitsu, W. Higemoto, R. Kadono, J. E. Sonier, A. N. Price, R. I. Miller, R. F. Kiefl, M. Nohara, H. Suzuki and H. Takagi: Phys. Rev. B {\bf 65} (2002) 140505.
\bibitem{Uemura:91} Y. J. Uemura, L. P. Le, G. M. Luke, B. J. Sternlieb, W. D. Wu, J. H. Brewer, T. M. Riseman, C. L. Seaman, M. B. Maple, M. Ishikawa, D. G. Hinks, J. D. Jorgensen, G. Saito and H. Yamochi: Phys. Rev. Lett. {\bf 66} (1991) 2665.
\bibitem{Takagiwa:05} H. Takagiwa E. Nishibori, N. Okada, M. Takata, M. Sakata and J. Akimitsu: unpublished.

\end{thebibliography}
\end{document}